\documentclass[final,3p,times, 12pt]{elsarticle}

\usepackage{graphics}
\usepackage{amsmath,amssymb}

\journal{Journal of Computational Physics}

\let\ph\varphi
\let\th\theta
\newcommand{\mw}{\mathcal{W}}
\newcommand{\mx}{\mathcal{X}}
\newcommand{\mA}{\mathcal{A}}
\newcommand{\mB}{\mathcal{B}}
\newcommand{\mC}{\mathcal{C}}
\newcommand{\tmb}{\tilde{\mathcal{B}}}
\newcommand{\tmc}{\hat{\mathcal{C}}}
\newcommand{\lapang}{\Delta_{\theta\varphi}}

\begin{document}

\begin{frontmatter}


\title{A spectral method for the wave equation of divergence-free vectors and
  symmetric tensors inside a sphere}

\author{J.~Novak\corref{cor1}}
\ead{jerome.novak@obspm.fr}
\cortext[cor1]{Corresponding author}

\author{J.-L.~Cornou}
\ead{jean-louis.cornou@obspm.fr}

\author{N.~Vasset}
\ead{nicolas.vasset@obspm.fr}

\address{Laboratoire Univers et Th\'eories, Observatoire de Paris, CNRS,
  Universit\'e Paris Diderot, 5 place Jules Janssen, F-92190, Meudon, France.}

\begin{abstract}
  The wave equation for vectors and symmetric tensors in spherical coordinates
  is studied under the divergence-free constraint. We describe a numerical
  method, based on the spectral decomposition of vector/tensor components onto
  spherical harmonics, that allows for the evolution of only those scalar
  fields which correspond to the divergence-free degrees of freedom of the
  vector/tensor. The full vector/tensor field is recovered at each time-step
  from these two (in the vector case), or three (symmetric tensor case) scalar
  fields, through the solution of a first-order system of ordinary
  differential equations (ODE) for each spherical harmonic. The correspondence
  with the poloidal-toroidal decomposition is shown for the vector
  case. Numerical tests are presented using an explicit Chebyshev-tau method
  for the radial coordinate.

\end{abstract}


\end{frontmatter}


\section{Introduction}
\label{s:intro}

Evolution partial differential equations (PDE) for vector fields under the
divergence-free constraint appear in many physical models. Similar problems
are to be solved with second-rank tensor fields. In most of these equations,
if the initial data and boundary conditions satisfy the divergence-free
condition, then the solution on a given time interval is divergence-free
too. But from the numerical point of view, things can be more complicated and
round-off errors can create undesired solutions, which may then trigger
growing unphysical modes. Therefore, in the case of vector fields, several
methods for the numerical solution of such PDEs have been devised, such as the
constraint transport method~\cite{evans-88} or the toroidal-poloidal
decomposition~\cite{dudley-89, marques-90}. The aim of this paper is to
present a new method for the case of symmetric tensor fields, which appear in
general relativity within the so-called 3+1 approach~\cite{alcubiere-08},
keeping in mind the vector case for which the method can be closely related to
the toroidal-poloidal approach. We first give motivations for the numerical
study of divergence-free vectors and tensors in Secs.~\ref{ss:GRMHD}
and~\ref{ss:numrel}; we briefly introduce our notations and conventions for
spherical coordinates and grid in Sec.~\ref{ss:spher}. The case of the vector
divergence-free evolution is studied in Sec.~\ref{s:vector}, and the link with
the poloidal-toroidal decomposition is detailed in Sec.~\ref{ss:pol_tor}. We
then turn to the symmetric tensor case in Sec.~\ref{s:sym_tensor} with the
particular traceless condition in Sec.~\ref{ss:TT_case}. A discussion of the
treatment of boundary conditions is given in Sec.~\ref{s:boundaries}, with the
particular point of inner boundary conditions
(Sec.~\ref{ss:excision}). Finally, some numerical experiments are reported in
Sec.~\ref{s:tests} to support our algorithms and concluding remarks are given
in Sec.~\ref{s:conc}.

\subsection{Divergence-free vector fields in relativistic magneto-hydrodynamics}
\label{ss:GRMHD}

In classical electro\-dynamics, the magnetic field is known to be
divergence-free since Max\-well's equa\-tions. This result can be extended to
general relativistic electrodynamics as well. In classical hydrodynamics, the
continuity equation can be expressed as $\partial_t \rho + \nabla \cdot (\rho
\mathbf{u}) = 0$, where $\rho$ is the mass density of the fluid, and
$\mathbf{u}$ its velocity. Various approximations give rise to divergence-free
vectors. Incompressible fluids have constant density along flow lines and therefore verify that
their velocity field $\mathbf{u}$ is divergence-free. Water is probably the most
common example of an incompressible fluid. In an astrophysical context, the
incompressible approximation can lead to a pretty good approximation of the
behavior of compressible fluid provided that the flow's Mach number is much
smaller than unity. Another useful hydrodynamic approximation is the anelastic
approximation, which essentially consists in filtering out the sound waves,
whose extremely short time scale would otherwise force the use of an
impractically small time step for numerical purposes. In general-relativistic
magneto-hydrodynamics, the anelastic approximation takes the form $\nabla
\cdot (\rho \Gamma \mathbf{u}) = 0$, where $\mathbf{u}$ is the coordinate
fluid velocity, $\Gamma$ the Lorentz factor of the fluid, and $\rho$ its
rest-mass density.

Divergence-free vectors have given rise to a large literature in numerical
simulations. For example, while using an induction equation to numerically
evolve a magnetic field, there is no guarantee that the divergence of the
updated magnetic field is numerically conserved. The most common methods to
conserve divergences in hyperbolic systems are constrained transport methods,
projection methods or hyperbolic divergence cleaning methods (see
\cite{toth-00} for a review).

\subsection{Divergence-free symmetric tensors in general relativity}
\label{ss:numrel}
The basic formalism of general relativity uses four-dimensional objects and,
in particular, symmetric four-tensors as the metric or the stress-energy
tensor. A choice of the gauge, which comes naturally to describe the
propagation of gravitational waves is the {\em harmonic gauge\/}
(e.g.~\cite{dedonder-21}), for which the divergence of the four-metric is
zero. The 3+1 formalism (see \cite{alcubiere-08} for a review) is an approach
to general relativity introducing a slicing of the four-dimensional
spacetime by three-dimensional spacelike surfaces, which have a Riemannian
induced three-metric. With this formalism, the four-dimensional tensors of
general relativity are projected onto these three-surfaces as
three-dimensional tensors. Consequently, the choice of the gauge on the
three-surface is a major issue for the computation of the solutions of
Einstein's equations. 

The divergence-free condition on the conformal three-metric has already been
put forward by Dirac~\cite{dirac-59} in Cartesian coordinates, and generalized
to any type of coordinates in~\cite{bonazzola-04}. This conformal three-metric
obeys an evolution equation which can be cast into a wave-like propagation
equation. Far from any strong source of gravitational field, this evolution
equation tends to a tensor wave equation, under the gauge constraint. With the
choice of the generalized Dirac gauge this translates into the system we study
in Sec.~\ref{s:sym_tensor}, with the addition of one extra constraint: the
fact that the determinant of the conformal metric must be one (Eq.~(167) of
\cite{bonazzola-04}). 

The choice of spherical coordinates and components comes naturally with the
study of isolated spheroidal objects as relativistic stars or black
holes. Moreover, boundary conditions for the metric or for the hydrodynamics
equations can be better expressed and implemented using tensor or vector
components in the spherical basis. The numerical simulations of
astrophysically relevant objects in general relativity must therefore be able
to deal with the evolution of divergence-free symmetric tensors, in spherical
coordinates and components. A particular care must be given to the fulfillment
of the divergence-free condition, since this additional constraint sets the
spatial gauge on the spacetime.

\subsection{Spherical components and coordinates}
\label{ss:spher}

In the following, unless specified, all the vector and tensor fields shall be
functions of the four spacetime coordinates $\mathbf{V}(t, r, \th, \ph )$ and
$\mathbf{h}(t, r, \th, \ph)$, where $(r, \th, \ph)$ are the polar spherical
coordinates. The associated spherical {\em orthonormal\/} basis is defined as:
\begin{equation}
\mathbf{e}_r = \frac{\partial}{\partial r}, \ 
\mathbf{e}_\th = \frac{1}{r}\frac{\partial}{\partial \th}, \ 
\mathbf{e}_\ph = \frac{1}{r\sin\th}\frac{\partial}{\partial \ph}.\label{e:spher_triad}
\end{equation}
The vector and symmetric tensor fields shall be described by their contravariant
components $\left\{V^r, V^\th, V^\ph\right\}$ and $\left\{ h^{rr}, h^{r\th},
  h^{r\ph}, h^{\th\th}, h^{\th\ph}, h^{\ph\ph} \right\}$, using this spherical basis:
\begin{equation}
  \label{e:def_components}
  \mathbf{V} = \sum_{i=r,\th,\ph} V^i\, \mathbf{e}_i,\quad \mathbf{h} =
  \sum_{i=r,\th,\ph}\sum_{j=r,\th,\ph} h^{ij}\, \mathbf{e}_i \otimes \mathbf{e}_j. 
\end{equation}
The scalar Laplace operator acting on a field $\phi(r, \th, \ph)$ is written:
\begin{equation}
  \label{e:scal_lap}
  \Delta \phi = \frac{\partial^2 \phi}{\partial r^2} + \frac{2}{r}
  \frac{\partial \phi}{\partial r} + \frac{1}{r^2} \lapang \phi,
\end{equation}
where $\lapang$ is the angular part of the Laplace operator, containing only
derivatives with respect to $\th$ or $\ph$:
\begin{equation}
  \label{e:lapang}
  \lapang \phi = \frac{\partial^2 \phi}{\partial \th^2} 
  + \frac{\cos\th}{\sin\th}\frac{\partial \phi}{\partial \th}
  + \frac{1}{\sin^2\th}
  \frac{\partial^2 \phi}{\partial \ph^2}.
\end{equation}

We now introduce scalar spherical harmonics, defined on the sphere as (see
Sec.~18.11 of \cite{boyd-01} for more details)
\begin{equation}
  \label{e:def_ylm}
  \forall \ell \geq 0,\ \forall m,\ 0 \leq m \leq \ell, \quad Y_\ell^m(\th,
  \ph) = e^{im\ph} P_\ell^m(\cos\th),
\end{equation}
where $P_\ell^m$ is the associated Legendre function. For negative $m$,
spherical harmonics are defined
\begin{equation}
  \forall m,\ -\ell \leq m < 0, \quad Y_\ell^m(\th, \ph) = (-1)^m e^{im\ph}
  P_\ell^{|m|}(\cos \th).
\end{equation}
Their two main properties used in this study are that they form a complete
basis for the development of regular scalar functions on the sphere, and that
they are eigenfunctions of the angular Laplace operator:
\begin{equation}
  \label{e:eigen_lapang}
  \forall (\ell, m), \ \lapang Y_\ell^m = -\ell(\ell+1) Y_\ell^m.
\end{equation}

\section{Vector case}
\label{s:vector}
We look for the solution of the following initial-boundary value problem of
unknown vector $\mathbf{V}$, inside a sphere of (constant) radius $R$, thus
$\forall (\th,\ph)$:
\begin{align}
  \forall t\geq 0,\ \forall r<R,\qquad& \frac{\partial^2 \mathbf{V}}{\partial
    t^2} =
  \mathbf{\Delta V},\label{e:vec_wave}\\
  \forall t\geq 0,\ \forall r\leq R,\qquad& \mathbf{\nabla}\cdot \mathbf{V} =
  0, \label{e:vec_div_condition}\\
  \forall r\leq R,\qquad& \mathbf{V}(0, r, \th, \ph) = \mathbf{v}_0(r, \th,
  \ph),\nonumber\\
  \forall r\leq R,\qquad& \left. \frac{\partial\mathbf{V}}{\partial t}
  \right|_{t=0}=
  \mathbf{w}_0(r, \th, \ph),\nonumber\\
  \forall t\geq 0,\qquad& \mathbf{V}(t, R, \th, \ph) = \mathbf{b}_0(t, \th,
  \ph).\label{e:BCvect}
\end{align} 
$\mathbf{v}_0, \mathbf{w}_0$ and $\mathbf{b}_0$ are given regular functions
for initial data and boundary conditions, respectively. $\mathbf{\Delta}$ is
the vector Laplace operator, which in spherical coordinates and in the
contravariant representation~(\ref{e:def_components}) using the orthonormal
basis~(\ref{e:spher_triad}) reads:
\begin{align}
  \label{e:vec_lap}
  \left( \mathbf{\Delta V} \right)^r ={}& \frac{\partial^2 V^r}{\partial r^2} +
  \frac{4}{r} \frac{\partial V^r}{\partial r} + \frac{2V^r}{r^2} +
  \frac{1}{r^2} \lapang V^r - \frac{2}{r} \Theta,\\
  \left( \mathbf{\Delta V} \right)^\th ={}& \frac{\partial^2 V^\th}{\partial
    r^2} + \frac{2}{r} \frac{\partial V^\th}{\partial r} + \frac{1}{r^2}
  \left( \lapang V^\th + 2 \frac{\partial V^r}{\partial \th} -
    \frac{V^\th}{\sin^2\th} - 2\frac{\cos\th}{\sin^2\th}\frac{\partial
      V^\ph}{\partial \ph}\right),\nonumber \\
  \left( \mathbf{\Delta V} \right)^\ph ={}& \frac{\partial^2 V^\ph}{\partial
    r^2} + \frac{2}{r} \frac{\partial V^\ph}{\partial r} + \frac{1}{r^2}
  \left( \lapang V^\ph + \frac{2}{\sin\th} \frac{\partial V^r}{\partial \ph} +
    2\frac{\cos\th}{\sin^2\th} \frac{\partial  V^\th}{\partial \ph} -
    \frac{V^\ph}{\sin^2\th} \right) \nonumber,
\end{align}
with the divergence $\Theta$
\begin{equation}
  \label{e:vec_div}
  \Theta \equiv \mathbf{\nabla}\cdot\mathbf{V} = \frac{\partial V^r}{\partial r} +
  \frac{2V^r}{r} + \frac{1}{r} \left( \frac{\partial V^\th}{\partial\th} +
    \frac{V^\th}{\tan\th} + \frac{1}{\sin\th}\frac{\partial V^\ph}{\partial
      \ph} \right).
\end{equation}
One can remark that a necessary condition for this system to be well-posed is
that 
\begin{equation}
  \label{e:vec_div_init}
  \mathbf{\nabla}\cdot \mathbf{v}_0 = \mathbf{\nabla}\cdot \mathbf{w}_0 = 0.
\end{equation}
In addition, the boundary setting at $r=R$ is actually overdetermined:
the three conditions are not independent because of the divergence
constraint. This aspect of the problem will be developed in more
details in Sec.~\ref{ss:bound_v}. 

In the rest of this Section, we devise a method to verify both equations
(\ref{e:vec_wave}) and (\ref{e:vec_div_condition}). This technique is similar
to that presented in~\cite{bonazzola-07} with the difference that we motivate
it by the use of vector spherical harmonics, and can easily be related to the
poloidal-toroidal decomposition method, as discussed in Sec.~\ref{ss:pol_tor}.

\subsection{Decomposition on vector spherical harmonics}
\label{ss:vec_ylm}

The first step is to decompose the angular dependence of the vector field
$\mathbf{V}$ onto a basis of {\em pure spin vector harmonics\/}
(see~\cite{thorne-80} for a review):
\begin{equation}
  \label{e:decomp_vec}
  \mathbf{V}(t, r, \th, \ph) = \sum_{\ell, m}  \left( E^{\ell m}(t, r)
    \mathbf{Y}^E_{\ell m} + B^{\ell m}(t, r) \mathbf{Y}^B_{\ell m} +
    R^{\ell m}(t, r) \mathbf{Y}^R_{\ell m} \right),
\end{equation}
defined from the scalar spherical harmonics as
\begin{align}
  \forall \ell > 0,\ \forall -\ell \leq m \leq \ell,\  \mathbf{Y}^E_{\ell m}
  ={}&  r\ \nabla Y_\ell^m,\\
  \forall \ell > 0,\ \forall -\ell \leq m \leq \ell,\  \mathbf{Y}^B_{\ell m}
  ={}&  \mathbf{e}_r \times \mathbf{Y}^E_{\ell m},\\
  \forall \ell \geq 0,\ \forall -\ell \leq m \leq \ell,\  \mathbf{Y}^R_{\ell
    m} ={}& Y_\ell^m\ \mathbf{e}_r;
\end{align}
where $\nabla$ is the gradient in the orthonormal
basis~(\ref{e:spher_triad}). Note that both $\mathbf{Y}^E_{\ell m}$ and
$\mathbf{Y}^B_{\ell m}$ are purely transverse, whereas $\mathbf{Y}^R_{\ell m}$
is purely radial. From this decomposition, we define the {\em pure spin\/}
components of $\mathbf{V}$ by summing all the multipoles with
{\em scalar\/} spherical harmonics~(\ref{e:def_ylm}):
\begin{align}
  V^\eta (t, r, \th, \ph) ={}& \sum_{\ell, m} E^{\ell m}\,
  Y_\ell^m, \label{e:def_etav}\\
  V^\mu (t, r, \th, \ph) ={}& \sum_{\ell, m} B^{\ell m}\,Y_\ell^m, \label{e:def_muv}
\end{align}
the last one being the usual $r$-component
\begin{equation}
  \label{e:def_vrv}
  \sum_{\ell, m} R^{\ell m}\,Y_\ell^m = V^r.
\end{equation}
The advantages of these pure spin components are first, that by construction
they can be expanded onto the scalar spherical harmonic basis, and second, that
angular derivatives appearing in all equations considered transform into the
angular Laplace operator~(\ref{e:eigen_lapang}).

To be more explicit, $(V^\eta, V^\mu)$ can be related to the vector spherical
components by (see also~\cite{bonazzola-04}):
\begin{align}
  V^{\th} = \frac{\partial V^\eta}{\partial \th} - \frac{1}{\sin \th}
  \frac{\partial V^\mu}{\partial \ph},   \label{e:etamudef_v} \\
  V^{\ph} = \frac{1}{\sin \th} \frac{\partial V^\eta}{\partial \ph} +
  \frac{\partial V^\mu}{\partial \th}; \nonumber
\end{align}
and inversely
\begin{align}
  \lapang V^\eta = \frac{\partial V^{\th}}{\partial \th} + \frac{V^{\th}}{\tan
    \th} + \frac{1}{\sin \th} \frac{\partial V^{\ph}}{\partial
    \ph}, \label{e:etasolve} \\ 
  \lapang V^\mu = \frac{\partial V^{\ph}}{\partial \th} + \frac{V^{\ph}}{\tan
    \th} - \frac{1}{\sin \th} \frac{\partial V^{\th}}{\partial \ph}. \label{e:musolve}
\end{align}
Let us here point out that the angular Laplace operator $\lapang$ is diagonal
with respect to the functional basis of spherical harmonics and, therefore,
the above relations can directly be used to obtain $V^\eta$ and $V^\mu$.

Thus, if the fields are defined on the whole sphere $\th \in [0, \pi],\ \ph
\in [0, 2\pi)$, it is possible to transform the usual components $\left(
  V^\th, V^\ph \right)$ to the pure spin ones $\left( V^\eta, V^\mu \right)$
by this one-to-one transformation, up to a constant ($\ell = 0$ part) for
$V^\eta$ and $V^\mu$. Since this constant is not relevant, it shall be set to
zero and disregarded in the following. Therefore, a vector field shall be
represented equivalently by its usual spherical components or by $\left( V^r,
  V^\eta, V^\mu\right)$.

\subsection{Divergence-free degrees of freedom}
\label{ss:vec_Amu}

From the vector spherical harmonic decomposition, we now compute two scalar
fields that represent the divergence-free degrees of freedom of a vector. We
start from the divergence of a general vector $\mathbf{W}$, expressed in terms
of pure spin components:
\begin{equation}
  \label{e:divfreevreta}
  \Theta = \frac{\partial W^r}{\partial r} + 2\frac{W^r}{r} + \frac{1}{r}\lapang W^\eta;
\end{equation}
where $W^\eta$ has been computed for the vector $\mathbf{W}$ from
Eq.~(\ref{e:etasolve}). This shows that the divergence of $\mathbf{W}$ does
not depend on the pure spin component $W^\mu$. On the other hand, it is
well-known that any sufficiently smooth and rapidly decaying vector field
$\mathbf{W}$ can be (uniquely on $\mathbb{R}^3$) decomposed as a sum of a
gradient and a divergence-free part (Helmholtz's theorem)
\begin{equation}
  \mathbf{W} = \nabla \phi + \mathbf{D}_0,
\end{equation}
with $\nabla \cdot \mathbf{D}_0 = 0$. From the formula~(\ref{e:musolve}), one
can check that the component $W^\mu$ only depends on $\mathbf{D}_0$. Next,
taking the curl of $\mathbf{W}$ and, in particular, combining the $\th$- and
$\ph$- components of this curl, one has that $\partial_r W^\eta +
\frac{W^\eta}{r} - \frac{W^r}{r}$ has the same property of being invariant
under the addition of any gradient field to $\mathbf{W}$, thus depends only on
$\mathbf{D}_0$. Therefore, we define the potential
\begin{equation}
  \label{e:defAv}
  A = \frac{\partial W^\eta}{\partial r} + \frac{W^\eta}{r} - \frac{W^r}{r}.
\end{equation}
 As a consequence, we have that
\begin{equation}
  \label{e:equiv_nulle}
  \mathbf{D}_0 = 0 \iff W^\mu = 0 \textrm{ and } A = 0.
\end{equation}
We have thus identified two scalar degrees of freedom for a divergence-free
vector field, which can be easily related to the well-known poloidal-toroidal
decomposition (Sec.~\ref{ss:pol_tor}), but have the advantage of being
generalizable to the symmetric tensor case.

We now write the wave equation~(\ref{e:vec_wave}) in terms of $V^\mu$ and $A$
(computed from $V^r$ and $V^\eta$). It is first interesting to examine the
pure spin components of the vector Laplace operator~(\ref{e:vec_lap}):
\begin{align}
  \left( \mathbf{\Delta V} \right)^\eta ={}& \Delta V^\eta + 2
  \frac{V^r}{r^2},\label{e:delta_v_eta} \\
  \left( \mathbf{\Delta V} \right)^\mu ={}& \Delta V^\mu; \label{e:delta_v_mu}
\end{align}
one sees that the equation for $V^\mu$ decouples from the system, therefore
Eq.~(\ref{e:vec_wave}) implies that
\begin{equation}
  \label{e:wave_muv}
  \frac{\partial^2 V^\mu}{\partial t^2} = \Delta V^\mu.
\end{equation}
Forming then from (\ref{e:vec_lap}) and~(\ref{e:delta_v_eta}) an equation for the
potential $A$, which is a consequence of the original wave
equation~(\ref{e:vec_wave}), we obtain 
\begin{equation}
  \label{e:wave_Av}
  \frac{\partial^2 A}{\partial t^2} = \Delta A.
\end{equation}
We are left with two scalar wave equations, (\ref{e:wave_muv}) and
(\ref{e:wave_Av}), for the divergence-free part of the vector field
$\mathbf{V}$. The recovery of the full vector field shall be discussed in
Sec.~\ref{ss:vec_scheme}; the treatment of boundary conditions shall be
presented in Sec.~\ref{ss:bound_v}.

\subsection{Link with poloidal-toroidal decomposition}
\label{ss:pol_tor}

According to the classical poloidal-toroidal decomposition, a divergence-free vector field $\mathbf{F}$ can be
considered to be generated by two scalar potentials $\Phi$ and $\Psi$, via 
\begin{equation}
  \label{e:poltordec} 
  \mathbf{F} = \mathbf{\nabla} \times (\Psi \mathbf{k}) + \nabla \times \nabla
  \times (\Phi \mathbf{k} ) 
\end{equation}
Here, $\mathbf{k}$ is a unit vector, called the pilot vector, which is chosen according to the geometry of the problem considered. In \cite{boronski-07a,
  boronski-07b}, $\mathbf{k}$ is chosen to be $\mathbf{e}_z$ in cylindrical
coordinates. One can also find the decomposition $\mathbf{F} = \mathbf{\nabla}
\times (A(r,\th)\mathbf{e_{\ph}}) + B(r,\th) \mathbf{e_{\ph}}$ when
considering axisymmetric solenoidal fields (see for example
\cite{Holler-02}). The latter representation makes $A$ appear clearly as a
poloidal component, and $B$ as a toroidal component. In order to link the
general poloidal-toroidal formalism to our previous potentials, we chose
$\mathbf{k} = \mathbf{e}_r$ in spherical coordinates (sometimes called the Mie
decomposition, see \cite{dormy-98} ). Then, one can show that
\begin{equation}
  \label{e:poltorcomp}
  \mathbf{F} = - \frac{1}{r^2}\lapang \Phi\, \mathbf{e}_r + 
  \frac{1}{r} \left( \frac{1}{\sin\th}\partial_{\ph} \Psi
    + \partial_{\th}\partial_r \Phi \right) \mathbf{e}_{\th} +  
  \frac{1}{r} \left( - \partial_{\th} \Psi +
    \frac{1}{\sin\th} \partial_{\ph} \partial_r \Phi \right) 
  \mathbf{e}_{\ph}
\end{equation}
Hence, we can identify the former pure spin components $F^\eta$ and $F^\mu$ through
\begin{align}
  \label{e:etamuphipsi}
  F^\eta = \frac{1}{r}\partial_r \Phi  \nonumber \\
  F^\mu = -\frac{1}{r} \Psi \nonumber
\end{align}
Therefore, the potential $A$ is linked to the potential $\Phi$ via
\begin{equation}
  \label{e:APhi}
  A = \frac{1}{r}\partial_r^2 \Phi + \frac{1}{r^3}\lapang\Phi = 
  \Delta \left(\frac{\Phi}{r} \right)
\end{equation}
which gives us a compatibility condition
\begin{equation}
  \label{e:compA}
  \lapang A = -\Delta(r F^r)
\end{equation}
The latter equation expresses that $\partial_r (r^2 \Theta) = 0$ for the original vector. Since our vector is a regular function of coordinates, it expresses that $\Theta = 0$. 

One can also show the following relations
\begin{align}
  \label{e:curlspoltor}
  \mathbf{e}_r \cdot \nabla \times \mathbf{F} = \frac{1}{r} \lapang F^\mu \nonumber \\
  \mathbf{e}_r \cdot \nabla \times \nabla \times \mathbf{F} = \frac{1}{r}
  \lapang A \nonumber 
\end{align}

\subsection{Integration scheme}
\label{ss:vec_scheme}

We defer to Sec.~\ref{ss:spher_SM} the numerical details about the integration
procedure, and we sketch here the various steps. From the result of
Sec.~\ref{ss:vec_Amu}, the problem
(\ref{e:vec_wave})-(\ref{e:vec_div_condition}) can be transformed into two
initial-value boundary problems, for the component $V^\mu$ (\ref{e:wave_muv})
and the potential $A$~(\ref{e:wave_Av}) respectively. Initial data can be
deduced from $\mathbf{v}_0$ and $\mathbf{w}_0$, so that $V^\mu(t=0)$ and
$\partial V^\mu / \partial t (t=0)$ are the $\mu$-components of, respectively,
$\mathbf{v}_0$ and $\mathbf{w}_0$. The same is true for the $A$ potential. The
determination of boundary conditions from the knowledge of $\mathbf{b}_0$
shall be discussed in Sec.~\ref{s:boundaries}. We therefore assume here that
we have computed the component $V^\mu$ and the potential $A$, inside the
sphere of radius $R$, for a given interval $[0, T]$, and we show how to
recover the whole vector $\mathbf{V}$.

The pure spin components $\left( V^r, V^\eta\right)$ of the vector $\mathbf{V}$
are obtained by solving the system of PDEs composed by the definition of the
potential $A$~(\ref{e:defAv}), together with the divergence-free
condition~(\ref{e:divfreevreta}). From their
definitions~(\ref{e:def_etav})-(\ref{e:def_vrv}), it is clear that the angular
parts of both $V^r$ and $V^\eta$ can be decomposed onto the basis of scalar
spherical harmonics, and therefore $A$ as well:
\begin{equation}
  \label{e:A_ylmv}
  A(t, r, \th, \ph) = \sum_{\ell, m} A^{\ell m}(t, r) Y_\ell^m(\th, \ph).
\end{equation}
We are left with the following set of systems of ordinary differential
equations in the $r$-coordinate:
\begin{equation}
\forall \ell > 0,\, \forall m\ -\ell \leq m \leq \ell,\quad \left\{
\begin{aligned}
\frac{\partial R^{\ell m}}{\partial r} + 2\frac{R^{\ell m}}{r} -
\frac{\ell(\ell+1)}{r} E^{\ell m} ={}& 0 \\
\frac{\partial E^{\ell m}}{\partial r} + \frac{E^{\ell m}}{r} - \frac{R^{\ell
      m}}{r} ={}& A^{\ell m}
\end{aligned}
\right.
.\label{e:sys_diracv}
\end{equation}
The potential $A$ being given, the pure spin components $V^r$ and $V^\eta$ are
obtained from this system, with the boundary conditions discussed in
Sec.~\ref{ss:bound_v}. The $\mu$-component is already known too, so it is
possible to compute the spherical components of $\mathbf{V}$ $\forall t\in [0,
T]$, from Eqs.~(\ref{e:etamudef_v}). Note that all angular derivatives present
in this system~(\ref{e:sys_diracv}) are only in the form of the angular
Laplace operator $\lapang$~(\ref{e:lapang}). It must also be emphasized that
the divergence-free condition is not enforced in terms of spherical components
(Eq.~(\ref{e:vec_div})), but in terms of pure spin components. Thus, if the
value of the divergence is numerically checked, it shall be higher than
machine precision, because of the numerical derivatives one must compute to
pass from pure spin to spherical components (Eqs.~(\ref{e:etamudef_v})).

The properties of the system~(\ref{e:sys_diracv}) are easy to
study. Substituting $R^{\ell m}$ in the first line by its expression as a
function of $E^{\ell m}$ and $A^{\ell m}$ (obtained from the second line), one
gets a simple Poisson equation:
\begin{equation}
  \Delta \left(rE^{\ell m}\right) = r \frac{\partial A^{\ell m}}{\partial r} +
  2A^{\ell m}.
\label{e:equPoisson}
\end{equation}
The discussion about boundary conditions, homogeneous solutions and regularity
for $r=0$ and $r\to \infty$ are immediately deduced from those of the Poisson
equation (see e.g.~\cite{grandclement-01}).

In the case where a source $\mathbf{S}$ is present on the right-hand side of
the problem~(\ref{e:vec_wave}), the method of imposing $\nabla \cdot
\mathbf{V} = 0$ can be generalized by adding sources to
Eqs.~(\ref{e:wave_muv})-(\ref{e:wave_Av}), which are deduced from
$\mathbf{S}$. Indeed, it is easy to show that the source for the equation for
$V^\mu$ is the pure spin $\mu$-component of $\mathbf{S}$ and the source for
the equation for $A$ is the equivalent potential computed from $\mathbf{S}$
pure spin components, using formula~(\ref{e:defAv}). Note that an
integrability condition for this problem is that the source be divergence-free
too. Therefore, for a well-posed problem, any gradient term present in
$\mathbf{S}$ can be considered as spurious and is naturally removed by this
method, since the $\mu$-component and the $A$ potential are both insensitive
to the gradient parts.

\section{Symmetric tensor case}
\label{s:sym_tensor}
Similarly to the vector case studied in Sec.~\ref{s:vector}, we look here for
the solution of an initial-boundary value problem of unknown symmetric tensor
$\mathbf{h}$, inside a sphere of radius $R$. As explained in
Sec.~\ref{ss:spher}, the symmetric tensor $\mathbf{h}$ shall be represented by
its contravariant components $h^{ij} (=h^{ji})$, where the indices run from
$1(r)$ to $3(\ph)$; moreover, we suppose that all components of $\mathbf{h}$
decay to zero at least as fast as $1/r$, as $r\to \infty$. We shall also use
the Einstein summation convention over repeated indices.

Thus the problem is written, $\forall (\th,\ph)$:
\begin{align}
  \forall t\geq 0,\ \forall r<R,\qquad& \frac{\partial^2 h^{ij}}{\partial t^2} =
  \Delta h^{ij},\label{e:tens_wave}\\ 
  \forall t\geq 0,\ \forall r\leq R,\qquad& \nabla_j  h^{ij} =
  0, \label{e:tens_div_condition}\\ 
  \forall r\leq R,\qquad& h^{ij}(0, r, \th, \ph) = \alpha^{ij}_0(r, \th, \ph),\nonumber\\
  \forall r\leq R,\qquad& \left. \frac{\partial h^{ij}}{\partial t} \right|_{t=0}=
  \gamma^{ij}_0(r, \th, \ph),\nonumber\\ 
  \forall t\geq 0,\qquad& h^{ij}(t, R, \th, \ph) = \beta^{ij}_0(t, \th, \ph).\label{e:BC_tens}
\end{align} 
The tensors $\alpha^{ij}_0, \gamma^{ij}_0$ and $\beta^{ij}_0$ are given
regular functions for initial data and boundary conditions, respectively. The
full expression of the tensor Laplace operator in spherical coordinates and in
the orthonormal spherical basis~(\ref{e:spher_triad}) is given by
Eqs.~(123)-(128) of~\cite{bonazzola-04} and shall not be recalled here. We
point out again that the boundary setting at $r=R$ is overdetermined: this is
discussed in more detail in Sec.~\ref{ss:bound_t}.

We introduce the vector $\mathbf{H}$, defined as the divergence of $h^{ij}$ and given
in the spherical contravariant components~(\ref{e:def_components}) by:
\begin{equation}
  H^i \equiv \nabla_j h^{ij} \iff \left\{
    \begin{aligned}
      H^r ={}& \frac{\partial h^{rr}}{\partial r} + \frac{2h^{rr}}{r} + \frac{1}{r}
      \left( \frac{\partial h^{r\th}}{\partial \th} + \frac{1}{\sin\th}
        \frac{\partial h^{r\ph}}{\partial \ph} - h^{\th\th} - h^{\ph\ph} +
        \frac{h^{r\th}}{\tan \th} \right), \\
      H^\th ={}& \frac{\partial h^{r\th}}{\partial r} + \frac{3h^{r\th}}{r} + \frac{1}{r}
      \left( \frac{\partial h^{\th\th}}{\partial \th} + \frac{1}{\sin \th}
        \frac{\partial h^{\th\ph}}{\partial \ph} + \frac{1}{\tan \th} \left(
          h^{\th\th} - h^{\ph\ph} \right) \right), \\
      H^\ph ={}& \frac{\partial h^{r\ph}}{\partial r} + \frac{3h^{r\ph}}{r} + \frac{1}{r}
      \left( \frac{\partial h^{\th\ph}}{\partial \th} + \frac{1}{\sin\th}
        \frac{\partial h^{\ph\ph}}{\partial \ph} + \frac{2 h^{\th\ph}}{\tan\th}
      \right) = 0.
    \end{aligned}
  \right. \label{e:def_H}
\end{equation}

We now detail, in the rest of this Section, a method to verify both evolution
equation (\ref{e:tens_wave}) and the divergence-free constraint
(\ref{e:tens_div_condition}).

\subsection{Decomposition on tensor spherical harmonics}
\label{ss:ten_ylm}

As in the vector case (Sec.~\ref{ss:vec_ylm}), we start by decomposing the
angular dependence of the tensor field $h^{ij}$ onto {\em pure spin tensor
  harmonics\/}, introduced by~\cite{mathews-62} and~\cite{zerilli-70} (we
again use the notations of~\cite{thorne-80}):
\begin{equation}
\mathbf{h}(t, r, \th, \ph) = \sum_{\ell, m} \left( L_0^{\ell m}
  \mathbf{T}^{L_0}_{\ell m}  + T_0^{\ell m} \mathbf{T}^{T_0}_{\ell m}  +
  E_1^{\ell m} \mathbf{T}^{E_1}_{\ell m}  + B_1^{\ell m}
  \mathbf{T}^{B_1}_{\ell m} + E_2^{\ell m} \mathbf{T}^{E_2}_{\ell m}  +
  B_2^{\ell m} \mathbf{T}^{B_2}_{\ell m} \right),
\label{e:decomp_ten}  
\end{equation}
where $\left( L_0^{\ell m}, T_0^{\ell m}, E_1^{\ell m}, B_1^{\ell m},
  E_2^{\ell m}, B_2^{\ell m} \right)$ are all functions of only
$(t,r)$. Complete definitions and properties of this set of tensor harmonics
can be found in~\cite{thorne-80}. Note that these harmonics have been devised
in order to describe gravitational radiation, far from any source. In that
respect, the most relevant harmonics are $\mathbf{T}^{E_2}$ and
$\mathbf{T}^{B_2}$, since they are transverse and traceless. The pure spin
components of the tensor $\mathbf{h}$ are defined as:
\begin{align}
  h^{rr} (t, r, \th, \ph) = \sum_{\ell, m} L_0^{\ell m}\,Y_\ell^m, \label{e:def_hrr}\\
  h^\tau (t, r, \th, \ph) = \sum_{\ell, m} T_0^{\ell m}\,Y_\ell^m, \label{e:def_tau}\\
  h^\eta (t, r, \th, \ph) = \sum_{\ell, m} E_1^{\ell m}\,Y_\ell^m, \label{e:def_eta}\\
  h^\mu (t, r, \th, \ph) = \sum_{\ell, m} B_1^{\ell m}\,Y_\ell^m, \label{e:def_mu}\\
  h^\mw (t, r, \th, \ph) = \sum_{\ell, m} E_2^{\ell m}\,Y_\ell^m, \label{e:def_w}\\
  h^\mx (t, r, \th, \ph) = \sum_{\ell, m} B_2^{\ell m}\,Y_\ell^m. \label{e:def_x}
\end{align}

Explicit relations between the last five components and the usual spherical
components~(\ref{e:def_components}) are now given.
\begin{equation}
  h^\tau = h^{\th\th} + h^{\ph\ph}
\end{equation}
is transverse; and the total trace is simply given by
\begin{equation}
  \label{e:def_trace}
  h = h^{rr} + h^\tau.
\end{equation}
In the following we shall use either the component $h^\tau$ or the trace. The
components $h^\eta$ and $h^\mu$ have similar formulas to those of the vector
pure spin components, as $\left\{ h^{ri} \right\}_{i=1, 2, 3}$ can be seen as
a vector:
\begin{align}
  h^{r\th} = \frac{\partial h^\eta}{\partial \th} - \frac{1}{\sin \th}
  \frac{\partial h^\mu}{\partial \ph},   \label{e:etamudef_t} \\
  h^{r\ph} = \frac{1}{\sin \th} \frac{\partial h^\eta}{\partial \ph} +
  \frac{\partial h^\mu}{\partial \th}; \nonumber
\end{align}
the reverse formula being similar to Eqs.~(\ref{e:etasolve})
and~(\ref{e:musolve}), they are not recalled here. Finally, the last two
components are obtained by:
\begin{align}
  P \equiv \frac{\left( h^{\th\th} - h^{\ph\ph} \right)}{2} ={}&
  \frac{\partial^2 h^\mw}{\partial \th^2} - \frac{1}{\tan \th} \frac{\partial
    h^\mw}{\partial \th} - \frac{1}{\sin^2 \th}
  \frac{\partial^2h^\mw}{\partial \ph^2} - 2 \frac{\partial}{\partial\th}
  \left( \frac{1}{\sin\th} \frac{\partial
      h^\mx}{\partial \ph}\right), \label{e:WXdef} \\
  h^{\th\ph} ={}& \frac{\partial^2 h^\mx}{\partial \th^2} - \frac{1}{\tan \th}
  \frac{\partial h^\mx}{\partial \th} - \frac{1}{\sin^2 \th}
  \frac{\partial^2h^\mx}{\partial \ph^2} + 2 \frac{\partial}{\partial\th}
  \left( \frac{1}{\sin\th} \frac{\partial h^\mw}{\partial \ph}
  \right);\nonumber
\end{align}
and the inverse relations are given by:
\begin{align}
  \lapang \left( \lapang + 2 \right) h^\mw ={}& \frac{\partial^2 P}{\partial
    \th^2} + \frac{3}{\tan \th} \frac{\partial P}{\partial \th} -
  \frac{1}{\sin^2 \th} \frac{\partial^2 P}{\partial \ph^2} -2P + \frac{2}{\sin
    \th} \frac{\partial}{\partial \ph} \left( \frac{\partial
      h^{\th\ph}}{\partial \th} +
    \frac{h^{\th\ph}}{\tan \th} \right) ,\label{e:compW}\\
  \lapang \left( \lapang + 2 \right) h^\mx ={}& \frac{\partial^2
    h^{\th\ph}}{\partial \th^2} + \frac{3}{\tan \th} \frac{\partial
    h^{\th\ph}}{\partial \th} - \frac{1}{\sin^2 \th} \frac{\partial^2
    h^{\th\ph}}{\partial \ph^2} -2h^{\th\ph} - \frac{2}{\sin \th}
  \frac{\partial}{\partial \ph} \left( \frac{\partial P}{\partial \th} +
    \frac{P}{\tan \th} \right) \label{e:compX}.
\end{align}
Here as for the vector case, the $h^\eta$ and $h^\mu$ components do not
contain any relevant $\ell = 0$ term, whereas $h^\mw$ and $h^\mx$ contain
neither $\ell = 0$, nor $\ell =1$ terms, as expected for transverse traceless
parts of the tensor $\mathbf{h}$. We shall use any set of components of the
tensor $\mathbf{h}$: either the usual ones $\left\{h^{ij}\right\}$, using the
spherical basis, or the pure spin ones $\left\{ h^{rr}, h^\tau \textrm{(or } h
  \textrm{)}, h^\eta, h^\mu, h^\mw, h^\mx \right\}$.

\subsection{Divergence-free degrees of freedom}
\label{ss:ten_AB}
The vector $\mathbf{H}$ defined as the divergence of $\mathbf{h}$ in
Eq.~(\ref{e:def_H}) can be expanded in terms of vector pure spin components,
which are then written as functions of the tensor pure spin components of
$\mathbf{h}$ (we use the trace $h$ instead of $h^\tau$):
\begin{align}
  H^r ={}& \frac{\partial h^{rr}}{\partial r} + \frac{3h^{rr}}{r} + \frac{1}{r}
  \left( \lapang h^\eta - h \right), \label{e:Hr}\\
  H^\eta ={}& \lapang \left[ \frac{\partial h^\eta}{\partial r} +
    \frac{3h^\eta}{r} + \frac{1}{r} \left( \left(\lapang + 2 \right)h^\mw +
      \frac{h - h^{rr}}{2} \right) \right], \label{e:Heta}\\
  H^\mu ={}& \lapang \left[ \frac{\partial h^\mu}{\partial r} +
    \frac{3h^\mu}{r} + \frac{1}{r} \left( \lapang + 2 \right) h^\mx \right]. \label{e:Hmu}
\end{align}
A possible generalization of the Helmholtz theorem to the symmetric tensor
case is that, for any sufficiently smooth and rapidly decaying symmetric
tensor field $\mathbf{T}$, one can find a unique (on $\mathbb{R}^3$)
decomposition of the form
\begin{equation}
  T^{ij} = \nabla^i L^j + \nabla^j L^i + h_0^{ij},
\end{equation}
with $\nabla_j h_0^{ij} = 0$. With these definitions, $\nabla_j T^{ij} = 0 \iff L^i
= 0$ which means that, from the six scalar degrees of freedom of the symmetric
tensor $T^{ij}$, the three longitudinal ones can be represented by the three
components of the vector $\mathbf{L}$. Therefore, the divergence-free
symmetric tensor $h_0^{ij}$ has only three scalar degrees of freedom that we
exhibit hereafter. 

One can check that the three scalar potentials defined by
\begin{align}
  \mA ={}& \frac{\partial T^\mx}{\partial r} -
  \frac{T^\mu}{r},\label{e:def_A}\\
  \mB ={}& \frac{\partial T^\mw}{\partial r} - \frac{1}{2r} \lapang
  T^\mw - \frac{T^\eta}{r} + \frac{T - T^{rr}}{4r}, \label{e:def_B}\\
  \mC ={}& \frac{\partial T}{\partial r} - \frac{\partial T^{rr}}{\partial
    r} + \frac{T}{r} - \frac{3T^{rr}}{r} - 2\lapang \left(
    \frac{\partial T^\mw}{\partial r} + \frac{T^\mw}{r}
  \right), \label{e:def_C}
\end{align}
satisfy the property
\begin{equation}
  \mA = \mB = \mC = 0 \iff \mathbf{h}_0 = 0,
\end{equation}
and represent the three divergence-free scalar degrees of freedom of a
symmetric tensor.

In order to write the wave equation~(\ref{e:tens_wave}) in terms of these
potentials, we first express the pure spin components of the tensor Laplace
operator acting on a general symmetric tensor $\mathbf{h}$: 
\begin{align}
  \left(\mathbf{\Delta h}\right)^{rr} ={}& \Delta h^{rr} - \frac{6h^{rr}}{r^2}
  -\frac{4}{r^2}\lapang h^\eta  + \frac{2h}{r^2} \label{e:dhrr} \\
  \left(\mathbf{\Delta h}\right)^\eta ={}& \Delta h^\eta + \frac{2}{r}\frac{\partial
    h^\eta}{\partial r} + \frac{2h^\eta}{r^2} - \frac{2}{r} \left(
    \frac{\partial h^\eta}{\partial r} + \frac{3h^\eta}{r} + \left( \lapang +
      2 \right) \frac{h^\mw}{r} + \frac{1}{2r}h - \frac{3}{2r} h^{rr}
  \right) \label{e:deta} , \\
  \left(\mathbf{\Delta h}\right)^\mu ={}& \Delta h^\mu + \frac{2}{r}\frac{\partial
    h^\mu}{\partial r} + \frac{2h^\mu}{r^2} - \frac{2}{r} \left( \frac{\partial
      h^\mu}{\partial r} + \frac{3h^\mu}{r} + \left(
      \lapang + 2 \right) \frac{h^\mx}{r} \right)\label{e:dmu} , \\
  \left(\mathbf{\Delta h}\right)^\mw ={}& \Delta h^\mw + \frac{2h^\mw}{r^2} +
  \frac{2h^\eta}{r^2},\label{e:dW} \\
  \left(\mathbf{\Delta h}\right)^\mx ={}& \Delta h^\mx + \frac{2h^\mx}{r^2} +
  \frac{2h^\mu}{r^2}, \label{e:dX}\\
  \textrm{trace of }\mathbf{\Delta h} ={}& \Delta h. \label{e:dT}
\end{align}
The term between parentheses in Eq.~(\ref{e:dmu}) is exactly zero in the case
of a divergence-free tensor, as it represents the $\mu$-component of the
vector $\mathbf{H}$~(\ref{e:Hmu}). The similar term in Eq.~(\ref{e:deta})
reduces to $-h^{rr}/r$, when using $H^\eta = 0$ with Eq.~(\ref{e:Heta}). We
can now write evolution equations, implied by the original tensor wave
equation~(\ref{e:tens_wave}):
\begin{align}
  \frac{\partial^2 \mA}{\partial t^2} ={}& \Delta
  \mA,\label{e:A_wave}\\
  \frac{\partial^2 \mB}{\partial t^2} ={}& \Delta \mB -
  \frac{\mC}{2r^2}, \label{e:B_wave}\\
  \frac{\partial^2 \mC}{\partial t^2} ={}& \Delta \mC +
  \frac{2\mC}{r^2} + \frac{8\lapang \mB}{r^2}. \label{e:C_wave}
\end{align}

The situation is therefore slightly more complicated than in the vector case
with Eqs.~(\ref{e:wave_muv})-(\ref{e:wave_Av}). Indeed, the two potentials
$\mB$ and $\mC$ are coupled, but it is possible to define new
potentials satisfying decoupled wave-like evolution equations. We first write
the scalar spherical harmonic decomposition of $\mA$, $\mB$ and
$\mC$:
\begin{align*}
  \mA(t, r, \th, \ph) ={}& \sum_{\ell, m} \mA^{\ell m}(t, r)
  Y_\ell^m(\th, \ph),\\
  \mB(t, r, \th, \ph) ={}& \sum_{\ell, m} \mB^{\ell m}(t, r)
  Y_\ell^m(\th, \ph),\\
  \mC(t, r, \th, \ph) ={}& \sum_{\ell, m} \mC^{\ell m}(t, r)
  Y_\ell^m(\th, \ph).
\end{align*}
Then, we define new potentials $\tmb$ and $\tmc$ as:
\begin{align}
  \tmb (t, r, \th, \ph) ={}& \sum_{\ell, m} \left( 2\mB^{\ell m}(t, r)
    + \frac{\mC^{\ell m}(t, r)}{2(\ell+1)} \right) Y_\ell^m(\th,
  \ph), \label{e:def_tB}\\
  \tmc (t, r, \th, \ph) ={}& \sum_{\ell, m} \left( \mC^{\ell m}(t, r)
      - 4\ell\mB^{\ell m}(t, r) \right) Y_\ell^m(\th, \ph). \label{e:def_tC}
\end{align}
The Eqs.~(\ref{e:B_wave})-(\ref{e:C_wave}) are transformed into:
\begin{align}
  \frac{\partial^2 \tmb}{\partial t^2} ={}& \tilde{\Delta} \tmb, \label{e:tB_wave}\\
  \frac{\partial^2 \tmc}{\partial t^2} ={}& \hat{\Delta} \tmc; \label{e:tC_wave}  
\end{align}
with, for any scalar field $f(r, \th, \ph) = \sum_{(\ell, m)} f^{\ell m}(r)
Y_\ell^m(\th, \ph)$, the operators defined as:
\begin{align}
  \tilde{\Delta} f ={}& \frac{\partial^2 f}{\partial r^2} + \frac{2}{r}
  \frac{\partial f}{\partial r} + \frac{1}{r^2} \left[ \sum_{\ell m}
    -\ell(\ell-1)f^{\ell m} Y_\ell^m \right], \label{e:def_tlap}\\
  \hat{\Delta} f ={}& \frac{\partial^2 f}{\partial r^2} + \frac{2}{r}
  \frac{\partial f}{\partial r} + \frac{1}{r^2} \left[ \sum_{\ell m}
    -(\ell+1)(\ell+2)f^{\ell m} Y_\ell^m \right]. \label{e:def_hlap}
\end{align}
These two operators are very similar to the usual Laplace operator, but in the
angular part $\lapang$, they contain a shift of, respectively $-1$ and $+1$ in
the multipolar number $\ell$, for $\tilde{\Delta}$ and $\hat{\Delta}$. We thus
have obtained three evolution wave-like
equations~(\ref{e:A_wave}), (\ref{e:tB_wave}) and (\ref{e:tC_wave}) for the
three scalar degrees of freedom of a divergence-free symmetric tensor.

\subsection{Traceless case}
\label{ss:TT_case}
As presented in Sec.~\ref{ss:numrel}, some evolution problems of symmetric
tensors in general relativity can have another constraint, in addition to the
divergence-free condition already studied~(\ref{e:tens_div_condition}). This
is the condition of determinant one for the conformal metric which turns into
an algebraic condition (Eq.~(169) of \cite{bonazzola-04}), and is enforced
by iteratively solving a Poisson equation with the trace of the tensor as a
source, as described in Sec.~V.D of \cite{bonazzola-04}. Therefore, in the
following the trace of the unknown tensor $\mathbf{h}$ is assumed to be known.

The fact that the trace $h$~(\ref{e:def_trace}) of a divergence-free symmetric
tensor is fixed reduces {\it a priori\/} the number of scalar degrees of
freedom to two. For instance, we here show that if the trace is given, the
scalar potentials $\mB$ and $\mC$ are linked. We take the
partial derivative with respect to $r$ of the definition of
$\mC$~(\ref{e:def_C}) and $\mB$~(\ref{e:def_B}) to obtain:
\begin{equation}
  \frac{\partial \mC}{\partial r} + \frac{2\mC}{r} + 2\lapang
  \left( \frac{\partial \mB}{\partial r} + \frac{3\mB}{r} -
    \frac{\mC}{4r} \right) = \Delta h.
\end{equation}
Therefore, if $h$ and $\mC$ are given, it is possible to integrate
this relation with respect to the $r$-coordinate to obtain $\mB$ (which we
have assumed to converge to $0$ as $r\to \infty$). Because of the
definitions~(\ref{e:def_tB})-(\ref{e:def_tC}), $\tmb$ and $\tmc$ are also
linked together if the trace is given.

We shall assume in the following that this trace is zero. All the equations
presented hereafter can easily be generalized to the non-zero (given) trace
case, taking the general form of the equations of Sec.~\ref{ss:ten_AB}. We
shall therefore use only two scalar potentials, namely $\mA$ and $\tmb$ to
describe a general traceless divergence-free symmetric tensor. 

\subsection{Integration scheme}
\label{ss:ten_evol}

Similarly to what has been done in the beginning of this section, we consider
the homogeneous wave equation for a symmetric tensor~(\ref{e:tens_wave}),
under the constraints that the tensor be
divergence-free~(\ref{e:tens_div_condition}) and traceless ($h = 0$). We have
seen in Sec.~\ref{ss:TT_case} that it was necessary to solve for at least the
two wave-like evolution equations~(\ref{e:A_wave}) and~(\ref{e:tB_wave}). We
describe now how to obtain the whole tensor, once $\mA(t, r, \th, \ph)$ and
$\tmb (t, r, \th, \ph)$ are known.

In order to obtain first the six pure spin components (actually, their
spherical harmonic decompositions~(\ref{e:def_hrr})-(\ref{e:def_x})) of
$\mathbf{h}$ at any time $t$, we use the following six equations: the
traceless condition, the three divergence-free conditions and the definitions
of $\mA$ and $\tmb$. They represent two systems of coupled differential
equations in the $r$-coordinate, that we express in terms of the tensor
spherical harmonic components~(\ref{e:decomp_ten}). The first one comes from
the definition of $\mA$~(\ref{e:def_A}) and the $H^\mu=0$
condition~(\ref{e:Hmu}); it couples the $\mu$- and the $\mx$-components of
$\mathbf{h}$:
\begin{align}
  &\frac{\partial B_2^{\ell m}}{\partial r} - \frac{B_1^{\ell m}}{r} =
  \mA^{\ell m}, \label{e:dirac_mag_1}\\
  &\frac{\partial B_1^{\ell m}}{\partial r} + \frac{3B_1^{\ell m}}{r} + \frac{2
    - \ell(\ell+1) B_2^{\ell m}}{r} = 0. \label{e:dirac_mag_2}
\end{align}
This system has two unknown functions $B_1^{\ell m}$ and $B_2^{\ell m}$,
whereas $\mA^{\ell m}$ is obtained from the time evolution of $\mA(t, r, \th,
\ph)$. 

The second one comes from the definition of $\tmb$~(\ref{e:def_tB}) and the
two $H^r=H^\eta=0$ conditions~(\ref{e:Hr})-(\ref{e:Heta}); it couples the
$rr$-, $\eta$- and $\mw$-components:
\begin{align}
  & (\ell + 2) \frac{\partial E_2^{\ell m}}{\partial r} + \ell(\ell + 2)
  \frac{E_2^{\ell m}}{r} - \frac{2 E_1^{\ell m}}{r} - \frac{1}{2(\ell + 1)}
  \frac{\partial L_0^{\ell m}}{\partial r} - \frac{\ell + 4}{\ell + 1}
  \frac{L_0^{\ell m}}{2r} = \tmb^{\ell m}, \label{e:dirac_elec_1}\\
  & \frac{\partial L_0^{\ell m}}{\partial r} + \frac{3 L_0^{\ell m}}{r} -
  \frac{\ell(\ell+1) E_1^{\ell m}}{r} = 0, \label{e:dirac_elec_2}\\
  & \frac{\partial E_1^{\ell m}}{\partial r} + \frac{3 E_1^{\ell m}}{r} -
  \frac{L_0^{\ell m}}{2r} + \frac{2 - \ell(\ell +1) E_2^{\ell m}}{r} =
  0. \label{e:dirac_elec_3} 
\end{align}
Here, the unknowns are $L_0^{\ell m}, E_1^{\ell m}$ and $E_2^{\ell m}$ and
$\tmb^{\ell m}$ is known from the evolution of $\tmb(t, r, \th, \ph)$.
 
When looking at a more general setting, the trace $h$ appears only in the
second system. If we combine Eq.~(\ref{e:dirac_mag_1}) with
Eq.~(\ref{e:dirac_mag_2}), we obtain a Poisson equation for the unknown
$rh^\mx$, with $\mA$ and its radial derivative as a source. As for the vector
case, this system can be solved using, for example, the spectral scalar
Poisson solver described in~\cite{grandclement-01}, and one obtains the pure
spin components $h^\mu$ and $h^\mx$. 

Such an argument cannot be used for the second system, but a search for
homogeneous solutions gives that, for a given $\ell$, the simple powers of
$r$:
\begin{equation}
  r^{\ell -2}, \frac{1}{r^{\ell+3}} \textrm{ and } \frac{1}{r^{\ell+1}}
\end{equation}
represent a basis of the kernel of the system
(\ref{e:dirac_elec_1})-(\ref{e:dirac_elec_3}). With this information, one can
devise a simple spectral method to solve this system (see
Sec.~\ref{ss:spher_SM}) and obtain the pure spin components $h^{rr}, h^\eta$
and $h^\mw$. With the traceless condition, one can also recover $h^\tau$ from
$h^{rr}$, and finally use Eqs.~(\ref{e:etamudef_t})-(\ref{e:WXdef}) to get the
spherical components of $\mathbf{h}$.

\section{Boundary conditions}
\label{s:boundaries}

\subsection{Vector system}
\label{ss:bound_v}

We discuss here the spatial boundary conditions to be used during our
procedure, so that we recover the unknown vector field at any time-step. The source of the vector wave equation is put to zero for the sake
of clarity; but the reasoning would be exactly the same in the general case.
 
As pointed out in Sec.~\ref{ss:vec_scheme}, the recovery of the vector field
at each time-step will require two different operations: first, we use the two
scalar wave equations~(\ref{e:wave_Av}) and~(\ref{e:wave_muv}) to recover $A$
and $V^{\mu}$. Two boundary conditions, set at the outer sphere (the boundary
of our computation domain), will then be needed for these quantities. The
second step will consist of the inversion of the differential
system~(\ref{e:sys_diracv}), to obtain the pure spin components $V^{r}$ and
$V^{\eta}$. This system is, in terms of the structure of the space
of homogeneous solutions,
mathematically equivalent to a Poisson problem (see Eq.~(\ref{e:equPoisson}));
its inversion will then also
require an additional boundary condition.
 
From the setting of our problem presented at the beginning of
Sec.~\ref{s:vector}, we can impose Dirichlet boundary conditions
for the 3 pure spin components on the outer sphere. The condition on $V^{\mu}$
enables us to recover the value of the entire field on our computational domain,
through the direct resolution of~(\ref{e:wave_muv}). Once we obtain the value
of the field $A$ on our domain, we can use a condition on either $V^{r}$ or
$V^{\eta}$ to invert the system~(\ref{e:sys_diracv}), and retrieve the
additional spin components.

There remains the necessity of imposing a boundary condition on $A$ to solve
Eq.~(\ref{e:wave_Av}). This cannot be done using condition at $r=R$
in~(\ref{e:BCvect}) and the definition~(\ref{e:defAv}), because
$\frac{\partial V^{\eta}}{\partial r}$ must be specified. To overcome this
difficulty, we exhibit here algebraic relations that link the value of $A$ at
the boundary and time derivatives of the pure spin components. These will be
compatibility conditions, derived only from the structure of our problem. We
express radial derivatives of equations~(\ref{e:divfreevreta})
and~(\ref{e:defAv}), respectively, to obtain, using relations~(\ref{e:vec_lap})
and~(\ref{e:delta_v_eta}), the following identities (see also
Eq.~(\ref{e:compA})):
\begin{eqnarray}
  \frac{1}{r} \lapang A = -\frac{\partial^{2} V^{r}}{\partial
    t^{2}}, \label{e:compAvec1}\\ 
  \frac{\partial A}{\partial r} + \frac{A}{r} = \frac{\partial^{2} V^{\eta}}{\partial
    t^{2}}, \label{e:compAvec2}.
\end{eqnarray}
 
Those equations are derived using only the fact that our vector field
satisfies the wave equation and is divergence-free. From the knowledge of the vector field at the boundary, we can impose either
of these two relations as boundary conditions for $A$; the first being of
Dirichlet type for each spherical harmonic of $A$, the second of Robin
type. This way we are able to solve equation~(\ref{e:wave_Av}), and complete
our resolution scheme.

Let us finally note that our boundary problem is, as one could guess, actually
overdetermined: there is no need to know the value of the entire vector field
on the outer sphere. It can be easily seen that, if one only has access to the
boundary values of $V^{\mu}$ and $V^{r}$, or $V^{\mu}$ and $V^{\eta}$, the
boundary conditions for all equations can be provided. This also gives us
insight about what would happen if we set up a numerical problem in which
spatial boundary conditions are not consistent with a solution of
Eqs.~(\ref{e:vec_wave},~\ref{e:vec_div_condition}); this could occur for
example because of numerical rounding errors or simply a physical boundary
prescription which is not compatible with a divergence-free vector field. Our
method will then still provide a solution that is divergence-free and which
satisfies Eqs.~(\ref{e:vec_wave},~\ref{e:vec_div_condition}); however only the
boundary conditions that are directly enforced will be satisfied. For example,
if we choose in our scheme to enforce boundary conditions on $V^{\mu}$ and
$V^{\eta}$, the outer boundary conditions that are satisfied at each time-step
are actually of the form (we keep the notation of (\ref{e:BCvect})):
\begin{eqnarray}
\forall t\geq 0,\qquad V^{\mu}(t, R, \th, \ph) &=& b_0^{\mu}(t, \th,
  \ph), \nonumber \\
V^{\eta}(t,R, \th, \ph) &=&b_0^{\eta}(t, \th,\ph), \nonumber \\
\frac{\partial V^{r}(t,R, \th, \ph)}{\partial r} + \frac{2}{r}
V^{r}(t,R, \th, \ph) &=& - \frac{1}{r}\lapang b_0^{\eta}(t, \th,\ph). \label{e:BCvect2}
\end{eqnarray}

The last condition is directly derived from the vanishing of
the divergence (Eq.~(\ref{e:divfreevreta})) at the boundary. Let us note that we do
not even impose a Dirichlet condition on $V^{r}$ as was originally intended. 
We may then not satisfy all the boundary conditions we
wished to prescribe at first. This may also depend on the boundary value we choose to use for the inversion of
the system~(\ref{e:sys_diracv}).

 We do not treat alternative
cases for the boundary problem (for which the knowledge of the vector field on
the outer sphere could be substituted by, for example, the knowledge of its
first radial derivative); but a similar approach would also provide
expressions for the boundary conditions of all the equations tackled in our
scheme.
  
\subsection{Tensor system}
\label{ss:bound_t}
The tensor problem presents itself in a similar way to the vector case,
only with a few additional difficulties. As seen in Sec.~\ref{ss:ten_evol}, we
can separate the problem into two parts; the first consists in retrieving the
field $\mathcal{A}$ from Eq~(\ref{e:A_wave}), and then get the spin components
$h^{\mu}$ and $h^{\mathcal{X}}$. In a similar way, we compute the value of
$\tilde{\mathcal{B}}$ from Eq.~(\ref{e:tB_wave}), so that we obtain the fields
$h^{rr}$, $h^{\mathcal{W}}$ and $h^{\eta}$ from the inversion of the
system~(\ref{e:dirac_elec_1},~\ref{e:dirac_elec_2},~\ref{e:dirac_elec_3}) .
The field $h^{\tau}$ is deduced from the traceless hypothesis. The tensor
field is then entirely determined.

As in the vector case, the solution of wave equations for $\mathcal{A}$ and
$\tilde{\mathcal{B}}$ requires one boundary condition for each equation. The
elliptic system~(\ref{e:dirac_mag_1},~\ref{e:dirac_mag_2}) is also quite similar
to that for the vector case, and its space of homogeneous solutions is also equivalent to that of a single Poisson equation. One boundary condition is
also required; it will be chosen as a Dirichlet condition on either $h^{\mu}$
or $h^{\mathcal{X}}$ , according to the setting of our
problem~(\ref{e:BC_tens}).

For the elliptic
system~(\ref{e:dirac_elec_1},~\ref{e:dirac_elec_2},~\ref{e:dirac_elec_3}), the
homogeneous solutions have been characterized in Sec.~\ref{ss:ten_evol}. The
only basis vector of the kernel of solutions that is regular in our
computation domain is, for any $\ell\geq 2$, the solution $r^{\ell -2}$. The
other two vectors of the kernel basis are not regular at the origin of
spherical coordinates. This means, from a basic point of view, that one
boundary condition will be sufficient at the outer sphere. It will be
provided, again according to our problem setting, as a Dirichlet condition on
any of the fields $h^{rr}$, $h^{\eta}$ or $h^{\mathcal{W}}$.

The last boundary problem concerns the fields $\mathcal{A}$ and
$\tilde{\mathcal{B}}$. They will be handled the same way as in the vector
case. We take the radial derivatives of the equations~(\ref{e:Hmu})
and~(\ref{e:def_A}), using the elliptic equations~(\ref{e:dmu})
and~(\ref{e:dX}), to obtain the following compatibility conditions:
\begin{eqnarray}
  (\lapang +2) \mathcal{A} = -\frac{\partial^{2} h^{\mu}}{\partial
    t^{2}}, \label{e:compAten1}\\ 
  \frac{\partial \mathcal{A}}{\partial r} + 2\frac{\mathcal{A}}{r} = \frac{\partial^{2} h^{\mathcal{X}}}{\partial
    t^{2}}, \label{e:compAten2}.
\end{eqnarray}

These are again derived using only the divergence-free property of the
vector field as well as the verification of the main wave equation. Using the known value of, respectively, $h^{\mu}$ and $h^{\mathcal{X}}$ at the
outer boundary, we obtain either a Dirichlet boundary condition for each
spherical harmonic from the first relation, or a Robin condition with the
second one. Again those identities have been obtained only from the
equations of our problem and the definitions of the variables we use.

Taking the same path for the second part of the problem, we express radial
derivatives of Eqs.~(\ref{e:Hr}), ~(\ref{e:Heta}), ~(\ref{e:def_B}) and
~(\ref{e:def_C}) to obtain respectively, and for each spherical harmonic, the
following relations:
\begin{eqnarray}
  \frac{\partial^{2} L_{0}^{\ell m}}{\partial t^{2}}
  &=&  -\frac{1}{(2 \ell +1)r} \left[ \frac{(\ell +1)(\ell +2)}{2}
    \tmc^{\ell m} - \ell(\ell +1)(\ell -1) \tmb^{\ell m} \right]
  \label{e:compBten1}\\ 
  \frac{\partial^{2} E_{1}^{\ell m}}{\partial
    t^{2}} &=& \frac{1}{(2 \ell +1)r}\left[(\ell +1)(\ell
    -1)\tmb^{\ell m} + \frac{\ell +2}{2}\tmc^{\ell m} \right] \label{e:compBten2}\\
  \frac{\partial^{2} E_{2}^{\ell m}}{\partial
    t^{2}} &=& \frac{1}{2 \ell +1} \left[\frac{(\ell
      +1)}{2}\frac{\partial \tmb^{\ell m}}{\partial r} -
    \frac{1}{4}\frac{\partial \tmc^{\ell m}}{\partial r} - \frac{(\ell
      +1)(\ell +2)}{2}\frac{\tmb^{\ell m}}{r} - \frac{\ell -3}{4}
    \frac{\tmc^{\ell m}}{r} \right] \label{e:compBten3}\\ 
  \frac{\partial^{2} (L_{0}^{\ell m} +T_{0}^{\ell m})} {\partial
    t^{2}} &=& \frac{1}{2 \ell +1} \Bigg[\frac{(\ell +1)(\ell
    +2)}{2}\frac{\partial \tmc^{\ell m}}{\partial r} - \ell(\ell
  +1)(\ell +2)\frac{\partial \tmb^{\ell m}}{\partial r} + \ell(\ell
  +1)(\ell -1)^{2}\frac{\tmb^{\ell m}}{r} \nonumber \\
  &+& \frac{1}{2}(\ell +1)\left[\ell(\ell -3) + \ell + 4\right]
  \frac{\tmc^{\ell m}}{r}  \Bigg]. \label{e:compBten4}
\end{eqnarray}

When expressing the vanishing of the trace, the last equation can be transformed into:
\begin{eqnarray}
 \frac{\partial^{2} E_{2}^{\ell m}}{\partial t^{2}} &=& 
 \frac{1}{2\ell(\ell +1)(2\ell +1)}\Bigg[(\ell +1)\frac{\partial
     \tmc^{\ell m}}{\partial r} + 2\ell(\ell +1)\frac{\partial
     \tmb^{\ell m}}{\partial r} +\frac{(\ell +1)(\ell +4)}{2}
   \frac{\tmc^{\ell m}}{r} \nonumber \\
  &-& \ell(\ell +1)(\ell -3) \frac{\tmb_{\ell
       m}}{r} \Bigg]. \label{e:compBten4mod}
\end{eqnarray}

Although those equations involve both the fields $\tilde{\mathcal{B}}$ and
$\hat{\mC}$, one can easily see that combining them can lead to
conditions on the field $\tilde{\mathcal{B}}$ only. For example, the
combination of~(\ref{e:compBten1}) and~(\ref{e:compBten2}) provides, for each
index $\ell$:
\begin{equation}
\tmb^{\ell m} = \frac{r}{(\ell +1)(\ell -1)} \Bigg[\frac{\partial^{2}
    L_{0}^{\ell m}}{\partial t^{2}} + (\ell +1)  \frac{\partial^{2}
    E_{1}^{\ell m}}{\partial t^{2}} \Bigg], \label{e:compBdiric}
\end{equation}
which is interpreted as a Dirichlet boundary condition for
$\tilde{\mathcal{B}}$. Robin boundary conditions can be obtained from the
combination of Eqs.~(\ref{e:compBten3}), ~(\ref{e:compBten4}), and
either~(\ref{e:compBten1}) or~(\ref{e:compBten2}). The tensor boundary
problem is then entirely solved; tests for some of the boundary conditions
derived here are presented in Sec.~\ref{s:tests}. Let us note again that this
problem is overdetermined: concerning the first system, the knowledge of
a Dirichlet condition on either only $h^{\mu}$, or only $h^{\mathcal{X}}$ suffices to
provide boundary conditions for $\mathcal{A}$ and the
system~(\ref{e:dirac_mag_1},~\ref{e:dirac_mag_2}). For the part of the
algorithm related to $\tilde{\mathcal{B}}$, we easily see that Dirichlet
conditions for any two of the spin components $h^{rr}$, $h^{\eta}$ and
$h^{\mathcal{W}}$ are sufficient to solve the boundary problem.

  We finally point out that, in the same fashion as in the vector case,
  if the value $\beta_{0}^{ij}$ imposed as a Dirichlet condition for
  the tensor at the outer boundary (Eq.~(\ref{e:BC_tens})) is not
  consistent with the system, the boundary
conditions actually imposed on our scheme will be slightly different:
only the Dirichlet conditions for the pure spin components that are
explicitly enforced will be satisfied. Other boundary values will only
express the coherence with respect to the fact that the solution is indeed divergence
free. As done in Sec.\ref{ss:bound_v}, it is possible to express other
boundary conditions enforced in practice by using the expression for
the tensor divergence $\mathbf{H}$ as a function of the pure spin components.

\subsection{Working in a shell: inner boundary conditions}
\label{ss:excision}
We say a few words here about the resolution of the tensorial problem when our
computation domain is no longer an entire sphere, but is instead bounded on the
interior at a finite coordinate radius $r=R_{\textrm{in}}>0$. We add in our
setting the condition that, $\forall (\th,\ph)$:
\begin{align}
  \forall t\geq 0,\qquad& h^{ij}(t, R_{\textrm{in}}, \th, \ph) =
  \zeta^{ij}_0(t, \th, \ph).\nonumber
\end{align}

Physical information is then also provided at the internal boundary
(this is, again, an overdetermined set of boundary conditions). This new
geometry will imply the need for two inner boundary conditions to be imposed
for the wave equations on $\mathcal{A}$ and in $\tilde{\mathcal{B}}$. These
are easily found using the results of the last section and the knowledge of
Dirichlet boundary conditions on the inner and outer sphere for all
components. The system~(\ref{e:dirac_mag_1},~\ref{e:dirac_mag_2}) also needs
an additional (inner) boundary condition, imposed on either $h^{\mu}$ or
$h^{\mathcal{X}}$. There is, however, a slight subtlety concerning the triple
system~(\ref{e:dirac_elec_1},~\ref{e:dirac_elec_2},~\ref{e:dirac_elec_3}). As
seen in Sec.~\ref{ss:ten_evol}, the kernel of solutions to this system is of
dimension 3, and since our computational domain no longer includes
$r=0$, all 3 basis vectors of this kernel are regular in our
domain. This means that 3 boundary conditions have to be imposed overall for
inverting this system (in contrast with the sphere case, where we only imposed
one). Those three conditions are imposed here on either $h^{rr}$, $h^{\eta}$
or $h^{\mathcal{W}}$ on each limit of the domain. We have {\it a
  priori} the freedom to choose which boundary conditions we want to
impose, and where to impose them; numerical experimentation would be required to
indicate whether or not there are preferable choices.

To conclude this section, we mention also the work of~\cite{VNJ09} where the
authors used the formalism presented in this paper to solve a tensor elliptic
equation that is part of a formulation of the Einstein equations. The resolution
was made on a 3-space excised by a sphere of fixed coordinate radius, where
the tensor equation possessed a weak singularity property
(see~\cite{HNW}). The boundary condition problem was treated a little bit
differently, as all boundary conditions imposed were either emanating from the
very structure of the problem, or were not needed at all. This is a
consequence of the particular behavior of that operator at the boundary; on
this setting for the domain geometry, one boundary condition was imposed to
invert the system in $h^{\mu}$ and $h^{\mathcal{X}}$, and two for the system
involving $h^{rr}$, $h^{\eta}$ and $h^{\mathcal{W}}$.

\section{Numerical tests}
\label{s:tests}

\subsection{Spectral methods in a sphere}
\label{ss:spher_SM}

The numerical schemes presented in previous sections have been implemented
using a multi-domain spectral method in spherical coordinates (see
e.g.~\cite{boyd-01, hesthaven-07}, for general presentations
and~\cite{grandclement-09} for a more detailed description in the case of
numerical relativity). We have used the \textsc{lorene} numerical
library~\cite{lorene}, with scalar fields decomposed onto a basis of Chebyshev
polynomials, in several domains, for the $r$-coordinate, Fourier series for
the $\ph$-coordinate and either Fourier or associated Legendre functions for
the $\th$-coordinate ($P_\ell^m(\cos\th)$, see Sec.~\ref{ss:spher}). This last
option is obviously needed by our algorithms, which strongly rely on spherical
harmonics decompositions and on the angular part of the Laplace operator
$\lapang$. The other basis of decomposition (Fourier) is quite useful for
computing angular derivatives $\partial / \partial \th$ and operators such as
$1/\sin\th$, appearing in e.g.~(\ref{e:etamudef_v})
or~(\ref{e:etamudef_t}). The coordinate singularity on the $z$-axis ($\th =0,
\pi$) is naturally handled by the spherical harmonic decomposition basis. We
cope with the coordinate singularity at the origin ($r=0$), using an even/odd
radial decomposition basis (only even/odd Chebyshev polynomials), depending on
the parity of the multipole $\ell$ (see~\cite{bonazzola-90} and Sec.~3.2
of~\cite{grandclement-09}). The complete regularity requirement would be that,
for each multipole $\ell$ the radial Taylor expansion of a regular function
should include only $r^p$ with $p\geq \ell$. We have found however that the
simpler parity prescription described above is in practice sufficient for the
study of the wave or Poisson equations performed here.

The wave equations~(\ref{e:wave_muv})-(\ref{e:wave_Av}) and
(\ref{e:A_wave})-(\ref{e:tB_wave}) are integrated numerically by writing them
as first-order systems:
\begin{equation}
  \frac{\partial^2 \phi}{\partial t^2} = \Delta \phi \iff \left\{
    \begin{aligned}
      \frac{\partial \phi}{\partial t} ={}& \psi,\\
      \frac{\partial \psi}{\partial t} ={}& \Delta \phi.
    \end{aligned}
\right.
\end{equation}
After discretization in the angular coordinates using spherical harmonics, we
then use a third-order Adams-Bashforth (explicit) time-stepping scheme with a
fixed time-step $dt$ and a Chebyshev-tau technique in the radial
coordinate. The differential systems for the computation of pure spin
components from the divergence-free degrees of freedom, as
system~(\ref{e:sys_diracv}) in the vector case, or
systems~(\ref{e:dirac_mag_1})-(\ref{e:dirac_elec_3}) in the tensor case, are
solved at every time-step in the Chebyshev coefficient space. A tau method is
used to match together the solutions across the domains, and to impose the
boundary conditions at $r=R$.

\subsection{Vector wave equation}
\label{ss:vec_wave}

\begin{figure}[t]
\centerline{\includegraphics[width=0.7\columnwidth]{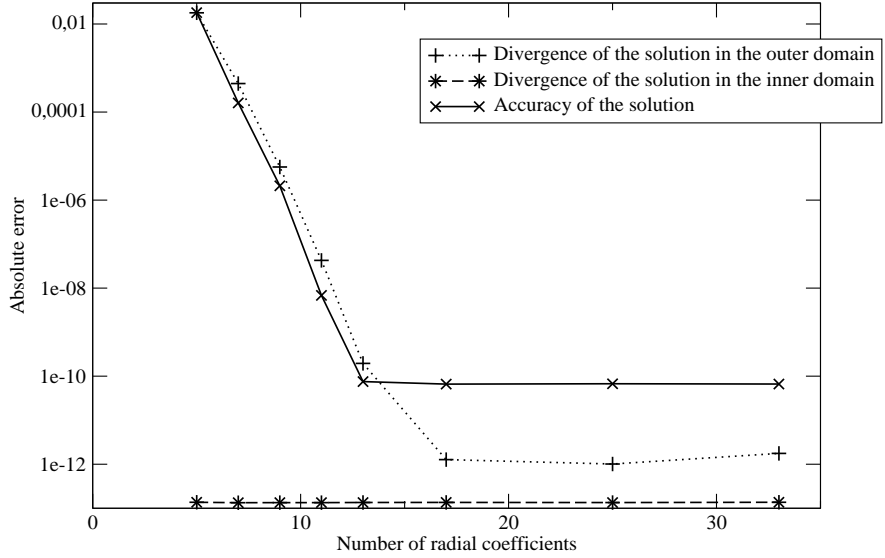}}
\caption[]{\label{f:CV_N_V} Decay of the errors (difference with theoretical
  solution and divergence of the numerical solution) for the vector wave
  equation, as a function of the number of radial Chebyshev coefficients $N_r$
  used in each domain. Other settings are $R=6, dt = 0,00032, N_\th=17,
  N_\ph=4$.}
\end{figure}

We consider here the numerical solution of the
problem~(\ref{e:vec_wave})-(\ref{e:BCvect}), with
$v_0^{i}(r, \th, \ph)$ given by its Cartesian components by (with $z=r\cos(\th)$):
\begin{equation}
  \label{e:ini_vect}
  v_0^{x} = -v_0^{y} = \cos(z),
\end{equation}
the other component is zero. Thus, the vector $v_0^{i}$ is clearly
divergence-free. With appropriate boundary conditions, the solution of the
problem~(\ref{e:vec_wave})-(\ref{e:BCvect}) is (still in Cartesian
components) simple to express:
\begin{equation}
  \label{e:sol_vect}
  V^{x}(t, r, \th, \ph) = - V^{y}(t, r, \th, \ph) = \cos(t)\cos(z),
\end{equation}
the other component being zero. The vector wave equation is solved through the
two scalar wave equations for the potentials $A$ and the component $V^\mu$ as
explained in Sec.~\ref{ss:vec_scheme}. From Eq.~(\ref{e:sol_vect}), we know
the values of $b_0^{i}(t, \th, \ph)$ appearing in Eq.~(\ref{e:BCvect}) as
Dirichlet boundary conditions and we can deduce its pure spin components
$\left( b_0^{r}, b_0^{\eta}, b_0^{\mu}\right)$. These are used to obtain
Dirichlet boundary conditions for the evolution equations for $A$ and $\mu$,
as described in Sec.\ref{ss:bound_v} using Eq.~(\ref{e:compAvec1}) for
$A$. Finally, the elliptic system~(\ref{e:sys_diracv}) is solved with the
appropriate Dirichlet boundary condition given by the spin component $b_0^{r}$
(see also Sec.~\ref{ss:bound_v}).

 \begin{figure}[t]
\centerline{\includegraphics[width=0.7\columnwidth]{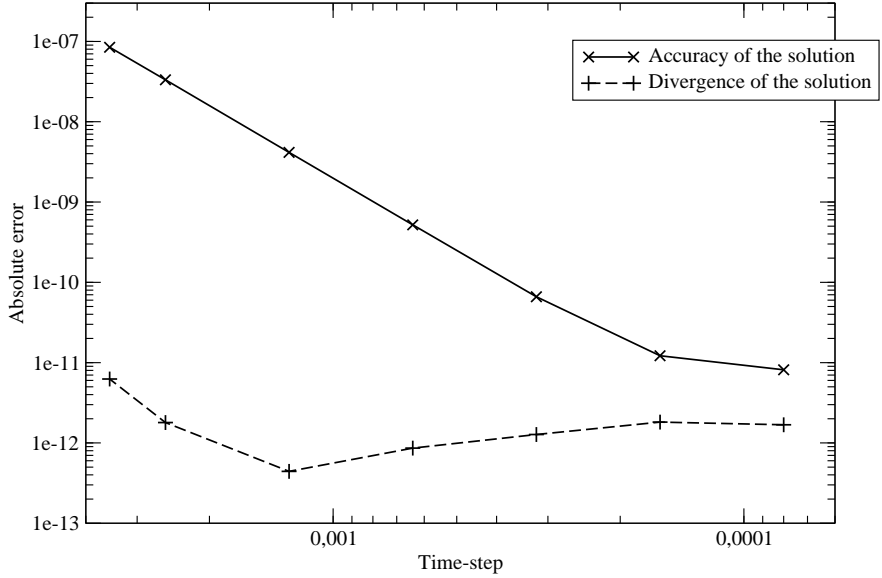}}
\caption[]{\label{f:CV_dt_v} Decay of the errors (difference with theoretical
  solution and divergence of the numerical solution) for the vector wave
  equation, as a function of the time-step $dt$. Other settings are $R=6,
  N_r=17, N_\th=17, N_\ph=4$.}
\end{figure}

We use the numerical techniques given in Sec.~\ref{ss:spher_SM}, with two
domains, and numbers of points in each direction given by $\left( N_r, N_\th,
  N_\ph \right)$. We have integrated the vector wave equation over the time
interval $t\in [0, 2\pi]$ and looked at the maximum in time of two quantities
to estimate the accuracy of the solution. First, the difference between the
numerical solution and the theoretical one~(\ref{e:sol_vect}), rotated to
spherical basis~(\ref{e:spher_triad}), is computed. Then, the divergence
of the numerical solution, expressed in the spherical basis is also
monitored. Note that, even though all the Cartesian components of $V^{i}$ do
not depend on the azimuthal angle $\ph$, the spherical components do depend on
$\ph$ and we have always used four points in the $\ph$-direction.

In Fig.~\ref{f:CV_N_V}, we observe as expected an exponential convergence of
both the discrepancy between the theoretical and numerical solutions (maximum
over all grid points and all components) as functions of the number of
spectral coefficients used in the radial direction $N_r$, all other parameters
being fixed. The same behavior has been observed when keeping $N_r$ fixed and
varying $N_\th$. Besides, we observe an exponential decay of the divergence of
the solution in the second (or outer) domain, whereas the divergence of the
solution in the first (central) domain remains constant to the radial
precision. This is due to the matching across domains and imposition of
boundary conditions, which can be seen as a modification of the solution of
the system~(\ref{e:sys_diracv}) by the addition of a linear combination of
homogeneous solutions. These homogeneous solutions of the system
(\ref{e:sys_diracv}) are, for each multipole $\ell$, $r^{\ell-1}$ and
$1/r^{\ell+2}$. The latter being singular at $r=0$ is not relevant in the
central domain. The $r^{\ell-1}$ function is a polynomial and is well
represented in the first domain, whereas in the second domain, we also need to
resolve $1/r^{\ell+2}$, which is poorly approximated for low values of
$N_r$. 

On the other hand, when varying the time-step $dt$, the difference between the
numerical and exact solutions decreases as $\mathcal{O}(dt^3)$ (see
Fig.~\ref{f:CV_dt_v}), as expected for a third-order scheme. Another feature
verified in Fig.~\ref{f:CV_dt_v} is the fact that the divergence of the
solution is (almost) independent of the time-step, being thus only a function
of the spatial resolution. The best accuracy observed in Fig.~\ref{f:CV_N_V}
is limited by angular resolution and the fact that the divergence is computed
using spherical components (Eq.~\ref{e:vec_div}), whereas the divergence-free
constraint is imposed using pure spin components
(Eq.~\ref{e:divfreevreta}). Therefore, the computation of derivatives in
Eqs.~(\ref{e:etamudef_v}) to obtain the spherical components introduces
additional numerical noise, depending on the angular resolution.

\subsection{Divergence-free and traceless tensor wave equation}
\label{ss:TT_wave}

\begin{figure}[t]
\centerline{\includegraphics[angle=-90,width=0.8\columnwidth]{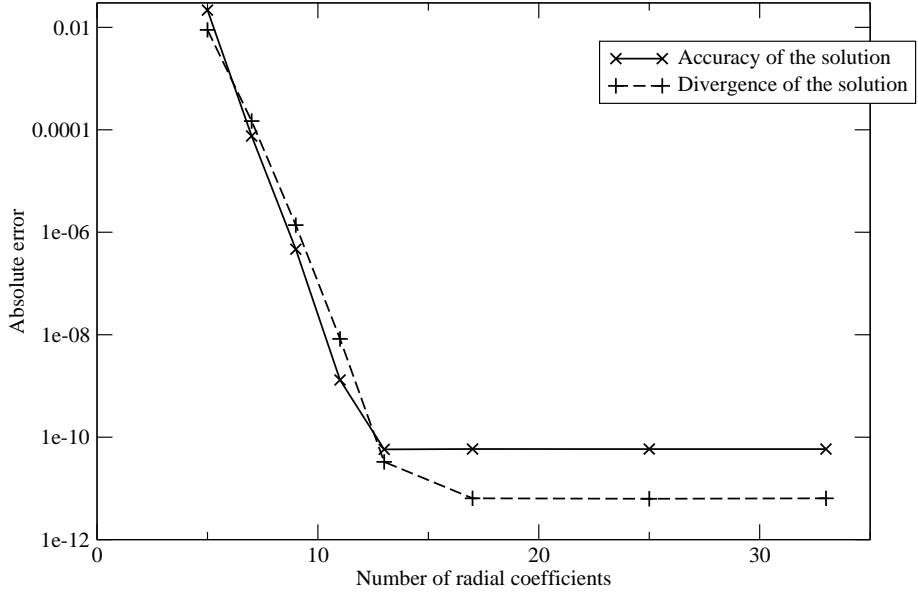}}
\caption[]{\label{f:CV_N_t} Decay of the errors (difference with theoretical
  solution and divergence of the numerical solution) for the tensor wave
  equation, as a function of the number of radial Chebyshev coefficients $N_r$
  used in each domain. Other settings are $R=6, dt=0.00032, N_\th=17,
  N_\ph=4$.}
\end{figure}

Similarly to Sec.~\ref{ss:vec_wave}, we consider here the numerical solution of the
problem~(\ref{e:tens_wave})-(\ref{e:BC_tens}), with
$\alpha_0^{ij}(r, \th, \ph)$ given in the Cartesian basis by (with $z=r\cos(\th)$):
\begin{equation}
  \label{e:ini_tens}
  \alpha_0^{xx} = -\alpha_0^{yy} = \cos(z),
\end{equation}
all the other components are zero. Thus the tensor $\alpha_0^{ij}$ is clearly
symmetric, divergence-free and trace-free. With $\gamma_0^{ij} = 0$ and
appropriate boundary conditions, the solution of the
problem~(\ref{e:tens_wave})-(\ref{e:BC_tens}) is (still in Cartesian
components) simple to express:
\begin{equation}
  \label{e:sol_tens}
  h^{xx}(t, r, \th, \ph) = - h^{yy}(t, r, \th, \ph) = \cos(t)\cos(z),
\end{equation}
all the other components being zero. The tensor wave equation is solved through
the two scalar wave-like equations for the potentials $\mA$ and $\tmb$ as
explained in Sec.~\ref{ss:ten_evol}. From Eq.~(\ref{e:sol_tens}), we know the
values of $\beta_0^{ij}(t, \th, \ph)$ appearing in Eq.~(\ref{e:BC_tens}) as
Dirichlet boundary conditions and we can deduce its pure spin components
$\left( \beta_0^{rr}, \beta_0^{\eta}, \beta_0^{\mu}\right)$. These are used to
obtain Dirichlet boundary conditions for the evolution equations for $\mA$ and
$\tmb$, as described in Sec.~\ref{ss:bound_t} using Eqs.~(\ref{e:compAten1})
and (\ref{e:compBdiric}), respectively. Finally, the elliptic
systems~(\ref{e:dirac_mag_1})-(\ref{e:dirac_elec_3}) are solved with the
appropriate Dirichlet boundary conditions given by the spin components of
$\beta_0^{ij}$, namely $\beta_0^{rr}$ and $\beta_o^\mu$. 
\begin{figure}[t]
\centerline{\includegraphics[angle=-90,width=0.8\columnwidth]{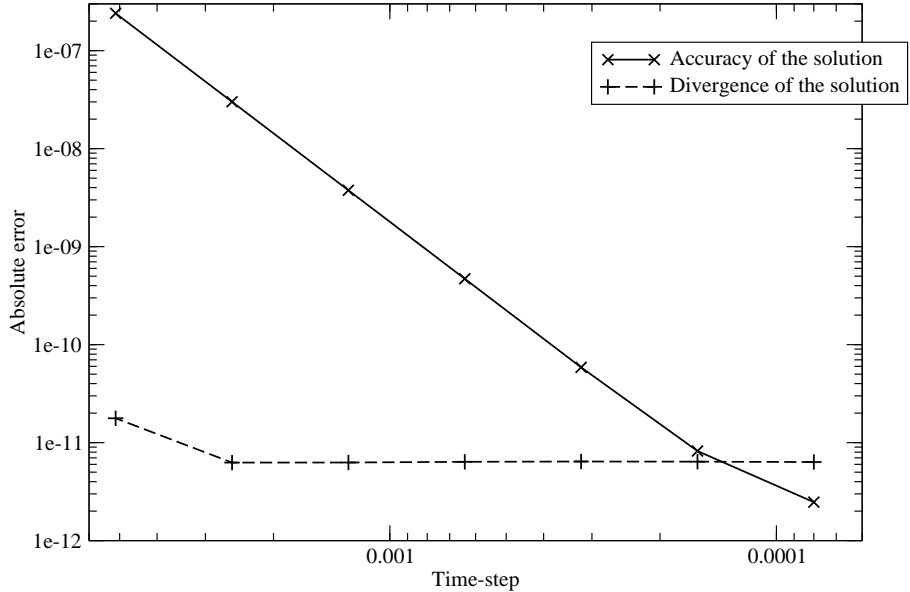}}
\caption[]{\label{f:CV_dt_t} Decay of the errors (difference with theoretical
  solution and divergence of the numerical solution) for the tensor wave
  equation, as a function of the time-step $dt$. Other settings are $R=6,
  N_r=17, N_\th=17, N_\ph=4$.}
\end{figure}
We have integrated the tensor wave equation following the same procedure as in
Sec.~\ref{ss:vec_wave}. results are displayed in Figs.~\ref{f:CV_N_t} and
\ref{f:CV_dt_t}, where we observe as expected an exponential convergence of
both the discrepancy between the theoretical and numerical solutions, and the
divergence of the numerical, as functions of $N_r$. When varying the time-step
$dt$, the difference between the numerical and exact solutions decreases as
$\mathcal{O}(dt^3)$, as expected. Here again, the divergence of the solution
is (almost) independent of the time-step, being thus only a function of the
spatial resolution, from the same reasons as in the vector case.

\section{Concluding remarks}
\label{s:conc}

We have described a new numerical method for solving the wave equation of a
rank-two symmetric tensor on a spherical grid, ensuring the divergence-free
condition on this tensor. In order to describe this method, we have first
addressed the vector case, for which we have reformulated the
poloidal-toroidal decomposition in spherical components. This approach, which
relies on a decomposition onto vector spherical harmonics was then generalized
to the case of a symmetric tensor. Through numerical tests of the vector and tensor
wave evolution in a sphere using spectral explicit time schemes, we have
observed that this method was convergent and accurate. In particular, the
level at which the divergence-free condition is violated is determined only by
the spatial discretization and does not depend on the time-step, as
expected. This method strongly relies on the decomposition onto spherical
harmonic spectral bases, but is not bound to spectral methods for the
representation of the radial coordinate.

The discussion in Sec.~\ref{s:boundaries} gave us the compatibility
conditions~(\ref{e:compAvec1}), (\ref{e:compAten1})~(\ref{e:compBdiric}),
which are necessary to obtain boundary conditions for the additional scalar
field equations, representing the evolution of the divergence-free degrees of
freedom of our objects ($A, \mA, \tmb$). The numerical tests performed in this
study have dealt only with simple Dirichlet boundary conditions. However, it
would be rather straightforward to generalize them to more complex boundary
conditions, which are needed in realistic simulations of gravitational
waves~\cite{lau-04, novak-04, rinne-09}.

In this respect, an interesting issue would probably be the general well-posed
nature of these boundary conditions with respect to our scheme, and how the
modifications for these conditions with this method, sketched in
Sec.\ref{ss:bound_v} and \ref{ss:bound_t}, would alter the physical behavior
of the solution. One could for example think of a Robin-like boundary setting
linked to an outer wave-absorbing condition (as in~\cite{novak-04}), instead
of the Dirichlet setting studied here; the fact that boundary conditions may
be only partially verified could have an effect on how this required feature
at the boundary would be described eventually in our scheme. The same type of
questions arise in a more general case, where the source terms of the
equations are non-vanishing: these sources would also require well-posedness
conditions (i.e. a vanishing divergence for the wave equation). If this
requirement is not satisfied (because of the iteration procedure or numerical
errors), although the problem is then mathematically ill-posed, our scheme
will still converge: it provides us with a solution of the wave equation with
a source that is basically the divergence-free part of the original ill-posed
source. The influence of this feature on the general stability and physical
relevance of the procedure is an open issue.

Future studies include the simulations of perturbed black hole spacetimes,
with the extraction of gravitational waves, and the solution of
general-relativistic magneto-hydrodynamics in the case of a rotating neutron
star.

\section*{Acknowledgments}
\label{sec:acknowledgements}

We wish to acknowledge the many fruitful discussions with Silvano Bonazzola
during the development of this method. We also thank Laurette Tuckerman for
critical reading of the manu\-script. This work was supported by the
A.N.R. Grants 06-2-134423 entitled ``M\'ethodes math\'e\-ma\-tiques pour la
relativit\'e g\'en\'erale'' and BLAN07-1\_201699 entitled ``LISA Science''.


\end{document}